\documentclass{ewic}
\usepackage{graphicx}

\begin{document}

\header{The \emph{Data Reduction and Analysis Graphical Organizer}}

\footer{\emph{Astronomical Data Analysis Conference III}}

\title{The \emph{Data Reduction and Analysis Graphical Organizer}}

\author{L. Paioro, B. M. Garilli, M. Scodeggio, P. Franzetti\\
Istituto Nazionale di Astrofisica\\
Istituto di Astrofisica Spaziale e Fisica cosmica - Sezione di Milano\\
via Bassini 15, I-20133 Milano, Italy\\
\email{luigi@mi.iasf.cnr.it}}

\begin{abstract}
Spectroscopic surveys are undergoing a rapid expansion in their data collecting
capabilities, reaching the level of hundreds of spectra per pointing.
An efficient use of such huge amounts of information requires a high degree of
interconnection between the various tools involved in preparing the
observations, reducing the data, and carrying out the data analysis.
DRAGO (Data Reduction and Analysis Graphical Organizer) attempts to easy the
process, by integrating in a global framework the main data handling components:
from reduction pipelines, to data organization, plotting, and browsing tools, to
storing the data reduction results in a database for further analysis.
DRAGO allows the use of the astronomer own's preferred tools, by "plugging them
in" in an environment which handles transparently the communications between
them. See http://cosmos.mi.iasf.cnr.it/pandora .
\end{abstract}

\keywords{Data Reduction, Data Organizer, Astronomical software, Database}

\section{INTRODUCTION}
The purpose of this package is to have a tool specifically built to handle
relatively large datasets, in the most complete way. The package must include
pipelines for data reduction to be carried out in a semi-automatic fashion,
tools for data visualization, and data analysis tools that can be used
efficiently for standard or ad-hoc analysis of the available data.

The need for such a package is dictated by the rapid increase in the number of
large telescopes available to the astronomical community, coupled with the
equally rapid increase in the multiplexing capabilities of the instruments
attached to those telescopes. While a normal long-slit spectrograph on a 4-meter
class telescope could produce a few tens of spectra per night of observation,
today a spectrograph like VIMOS at the VLT can obtain several thousands spectra
per night. This productivity increase has rendered obsolete traditional methods
of data reduction and analysis, at least as long as these data must be reduced
and analyzed in a timely fashion. It is clearly necessary to automatize as much 
as possible these operations, to increase the speed with which they can be
carried out,  but without sacrificing the capability of analyzing in detail the 
results of the various operations, and eventually intervene manually to change
the way some of these operations are carried out. Moreover it is necessary to
develop an efficient and rigorous data organizer and archiver, so that the
available data and files would not be lost among hundreds or thousands of 
similar data and files.

This concurrent need for automation of the data reduction and analysis
procedures, and for an efficient and rigorous data organizer leads to the
decision of building a new package, specifically designed to satisfy these 
needs. General-purpose astronomical software packages like IRAF or MIDAS in fact 
are not well equipped for these tasks, and to add these facilities using the
rather limited programming tools these packages offer would require a
prohibitively large amount of work.

\section{TECHNICAL DESCRIPTION}

We can identify three main types of operation: \\
- data organization (raw data and reduced data); \\
- data reduction and analysis; \\
- data storing (coming from data reduction and analysis).

Each of the tasks above corresponds to one (or more) software module.
Thus we have a File Browser for the data handling and managing, one 
or more Reduction Software for data elaboration and analysis, and a Database
Browser for the data storing and inquiring.

\begin{figure*}
  \begin{tabular}{c}
    \includegraphics[width=14.0cm]{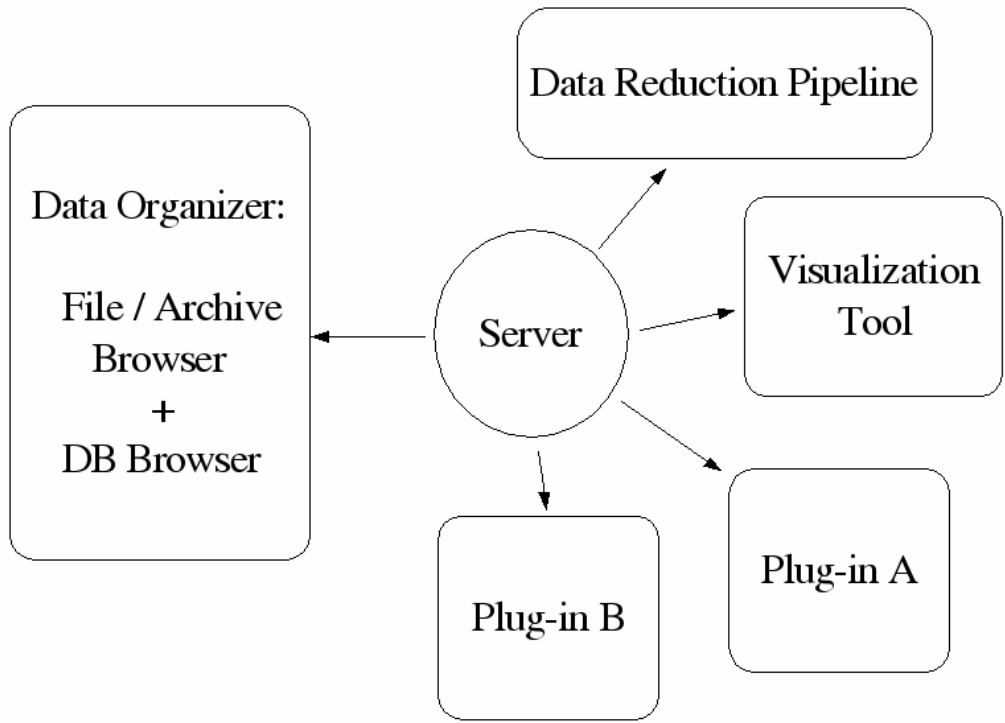}
  \end{tabular}
  \caption{Communication Server scheme. The Server sorts out the informations usign
   a communication protocol studied for plug-ins purpose.}
  \label{figschema}
\end{figure*}

Operationally, DRAGO is structured around a Communication Server, which
handles the communication between these independent software modules (see Figure ~\ref{figschema}).
Every functionality the package offers, is included in one such module,
and it is possible to add external plug-in modules following a very simple 
procedure.

\begin{figure*}
  \begin{tabular}{cc}
    \includegraphics[width=4.0cm]{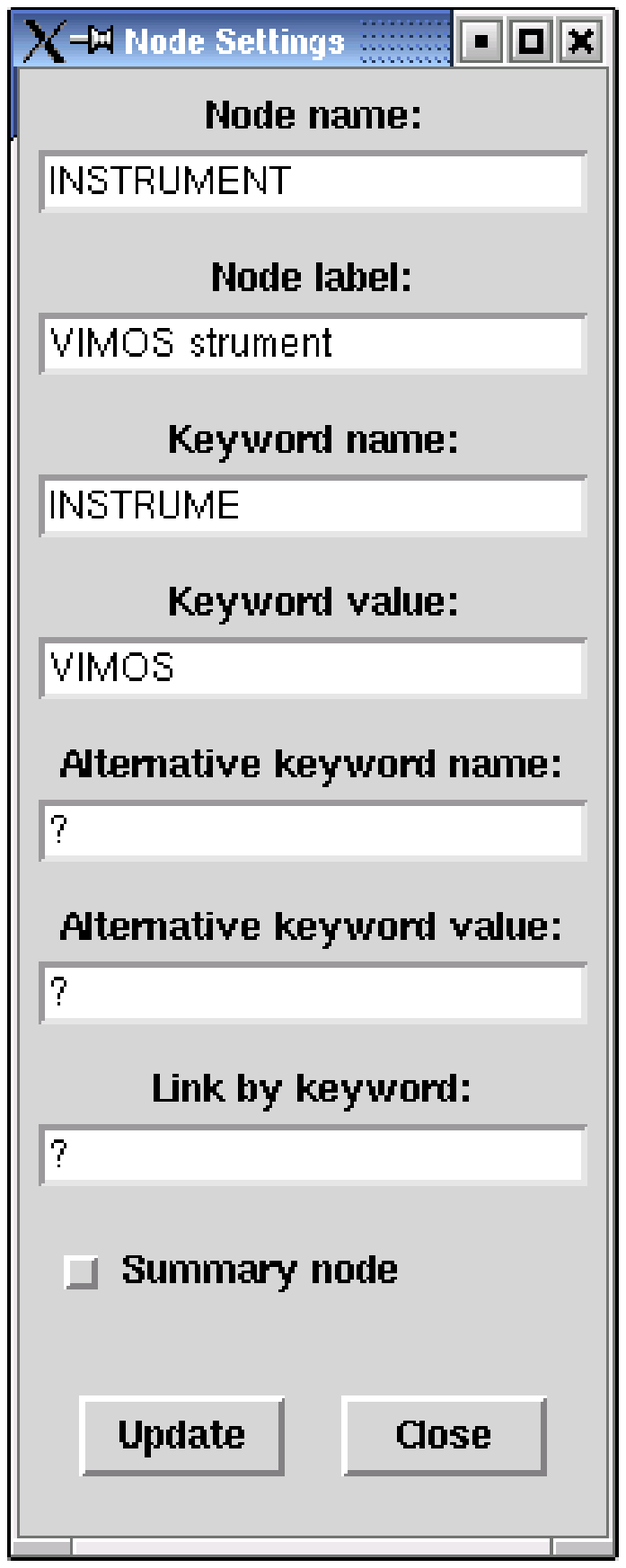}&
    \includegraphics[width=9.0cm]{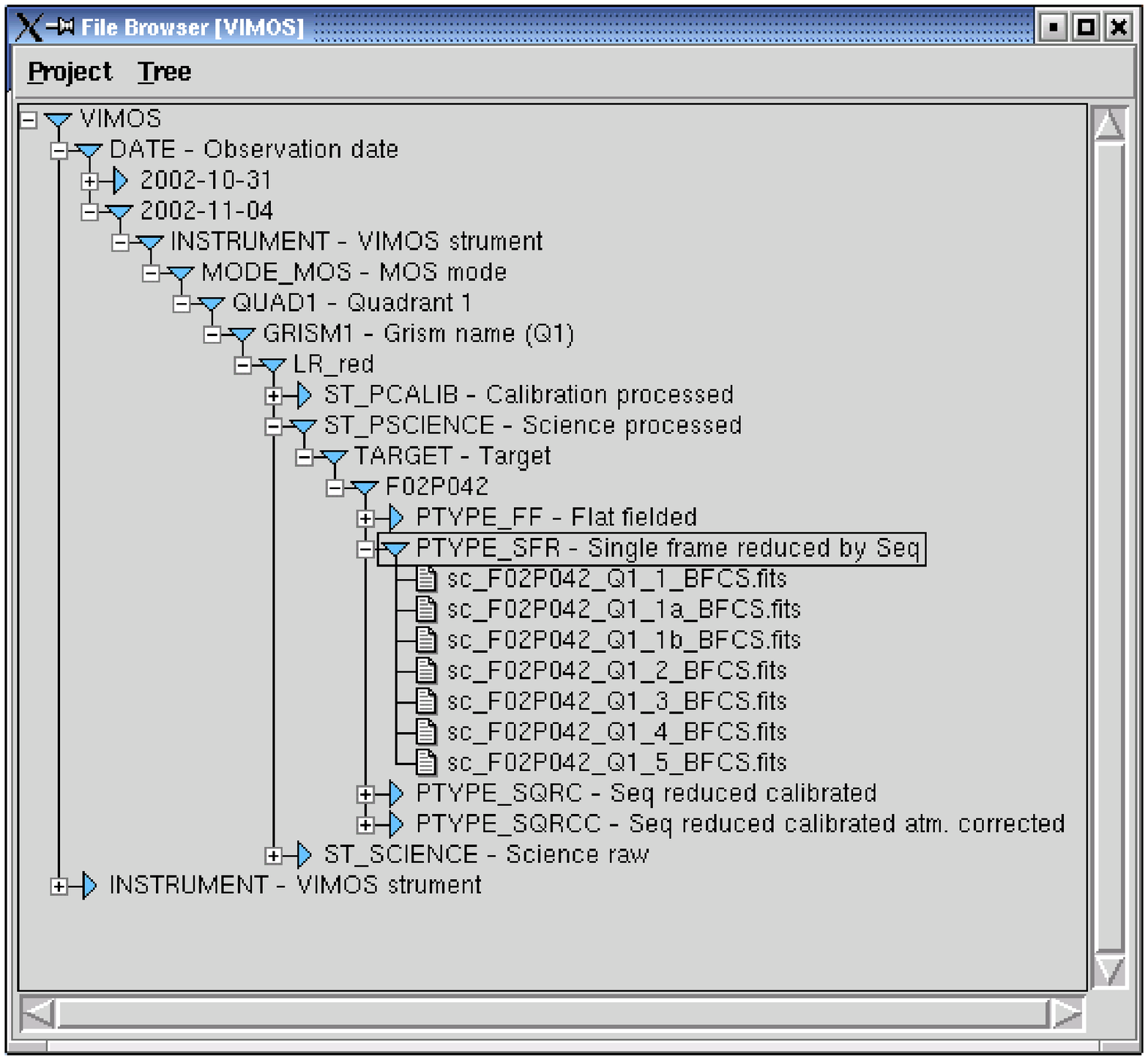}
  \end{tabular}
  \caption{Left: an user node setting example based on VIMOS one.
           Right: the File Browser used with the standard VIMOS tree.}
  \label{figfileb}
\end{figure*}

At the first development stage, the Communication Server is a relatively
simple one, and all package modules are local packages, so that all instances of all software tools
will be executed on a single workstation. In a future implementation, the communication
server will be replaced with an advanced server, capable of handling remote
communications and of taking advantage of distributed computing infrastructures,
where separate data reduction or data analysis modules and the data storage
facilities can be located anywere within a grid of computers.

The core DRAGO functions are being developed using Python. Data reduction and analysis plug-ins
can be coded using any language of choice, and they will require a small Python interface
to obtain a complete integration within the DRAGO package.

\section{FILE BROWSER}

The File Browser is responsible for data organization: it categorizes files according to
user's defined criteria, and allows an easy and fast browsing through them.

As an example, in Figure ~\ref{figfileb} we show the characterization of VIMOS data files:
the panel on the left is to be filled in by the user, and allows to define the data
structure on the basis of FITS keywords. On the right we show the categorization of data files
resulting from such definition.

\begin{figure*}
  \begin{tabular}{c}
    \includegraphics[width=14.0cm]{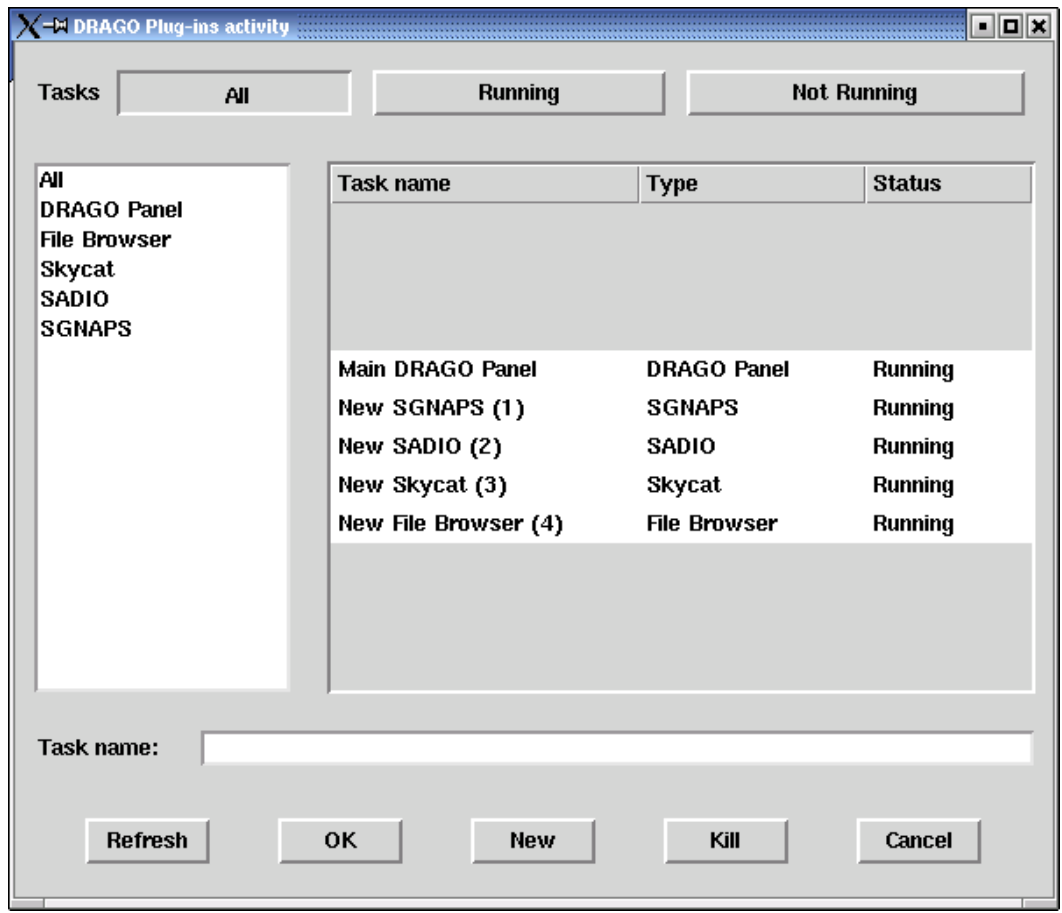}
  \end{tabular}
  \caption{The plug-ins manager. It handles the data exchange between plug-ins, sending
   informations through a Communication Server.}
  \label{figplugin}
\end{figure*}

\section{EXTERNAL PLUG-INS}

DRAGO provides a plug-in system where each
data reduction software and tool (coming from third parties) can be integrated in the global
framework. The astronomer can use his own's preferred tool exploiting
at the same time all of the others DRAGO potentialities.

DRAGO plug-in technology makes it extremely simple to incorporate existing software
tools, interfacing them in a single working framework.

To demostrate and test DRAGO plug-ins system, we have used it to plug-in a number of packages for data
visualization and analysis (see Figure ~\ref{figplugin}). Skycat is used as example of distributed
external software, while SADIO and SGNAPS are two internally developed packages we have incorporated as
testing case.

\begin{figure*}
  \begin{tabular}{cc}
    \includegraphics[width=7.0cm]{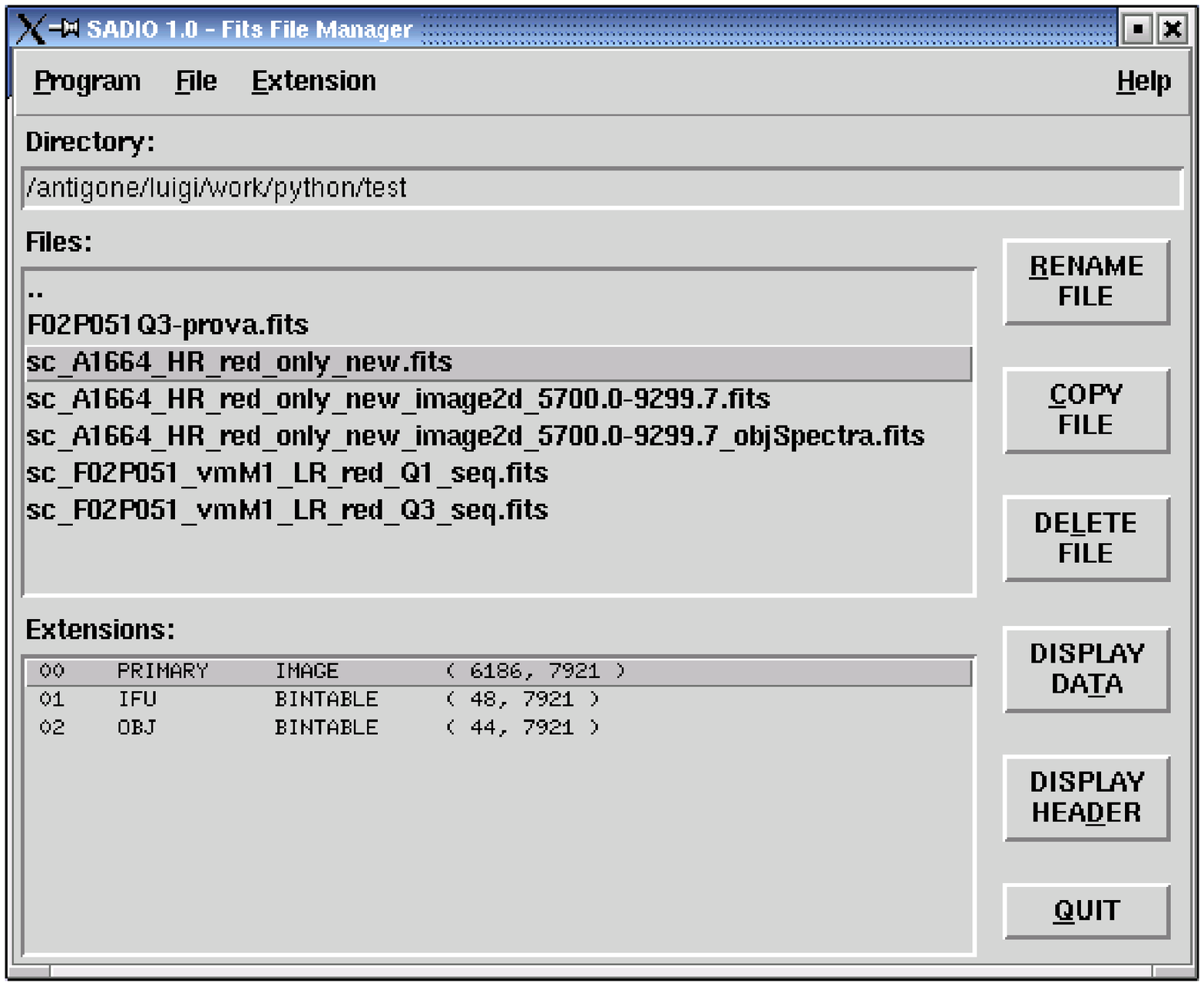} &
    \includegraphics[width=7.0cm]{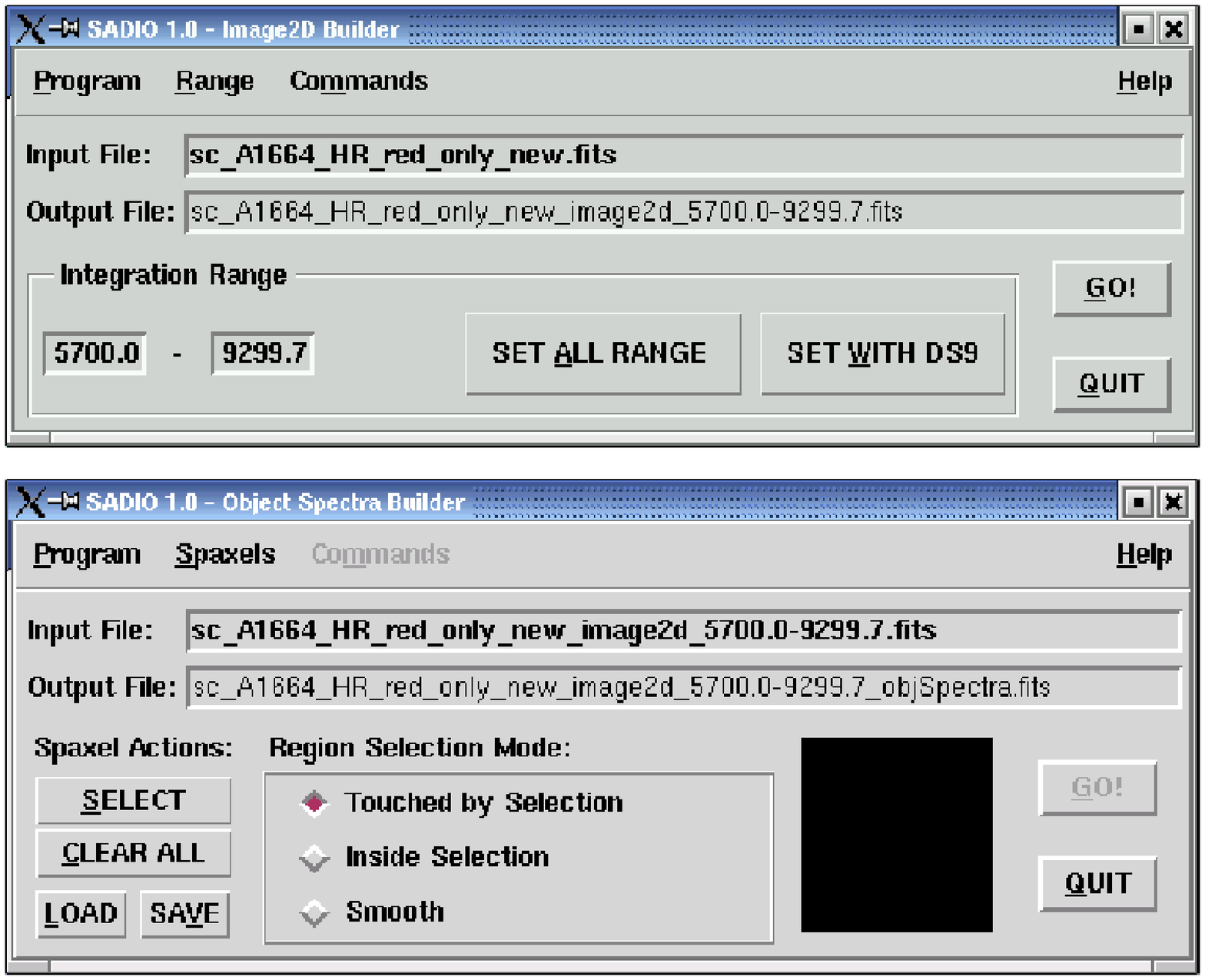}
  \end{tabular}
  \caption{SADIO plug-in. We have three screenshots with the File Manager (that
  allows to view the file content), the Image2D Builder (that allows to integrate
  over a wavelength range to create a 2D image) and the Object Spectra Builder (that creates
  a monodimensional spectrum based on a spaxel selection).}
  \label{figsadio}
\end{figure*}

SADIO plug-in (Software for Analise and Display IFU Observations) allows to elaborate
VIMOS-IFU data. It can produce a 2D image
integratig spectra over a user defined wavelength range, and a monodimensional spectrum, spatially
integrating spectra over a region selected from the 2D image produced before (Figure ~\ref{figsadio}).

\begin{figure*}
  \begin{tabular}{c}
    \includegraphics[width=14.0cm]{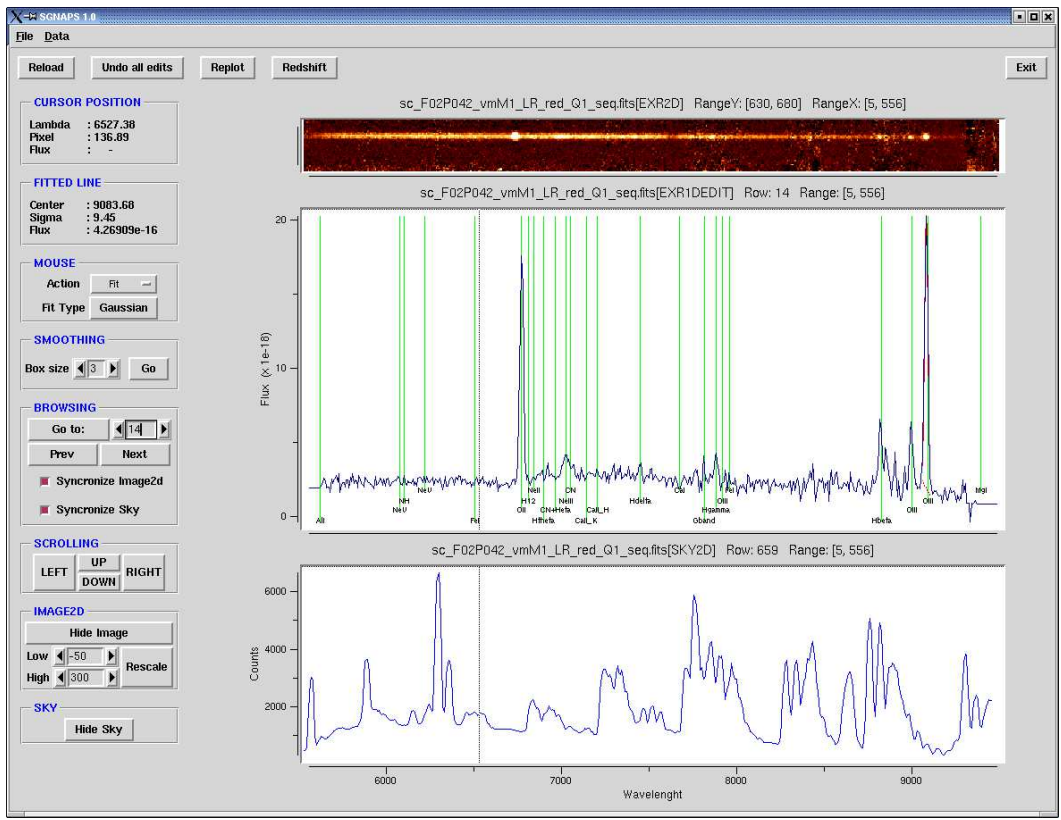}
  \end{tabular}
  \caption{SGNAPS plug-in. An example of redshift analysis on a VIMOS spectrum.}
  \label{figsgnaps}
\end{figure*}

SGNAPS (Software for Graphical Navigation, Analysis and Plotting of Spectra)
allows to display and elaborate 1D spectra, coordinating its display with that of 
the corresponding 2D image and sky spectrum. It also provides a  redshift estimation tool
(Figure ~\ref{figsgnaps}).

\section{DATABASE}

The last plug-in we will provide within DRAGO is an easy-to-use interface to MySQL named DBrowser.

\begin{figure*}
  \begin{tabular}{cc}
    \includegraphics[width=7.0cm]{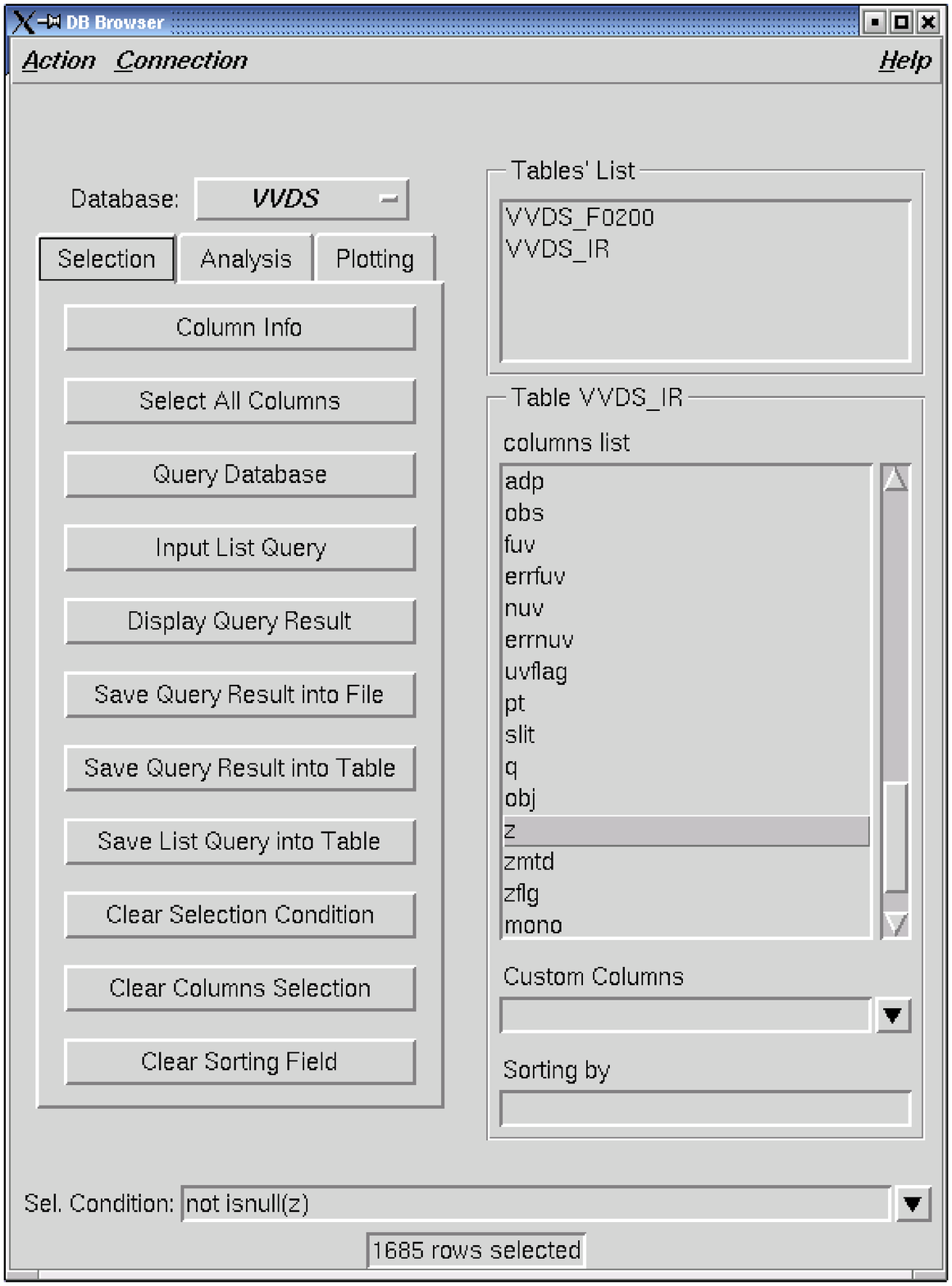}&
    \includegraphics[width=7.0cm]{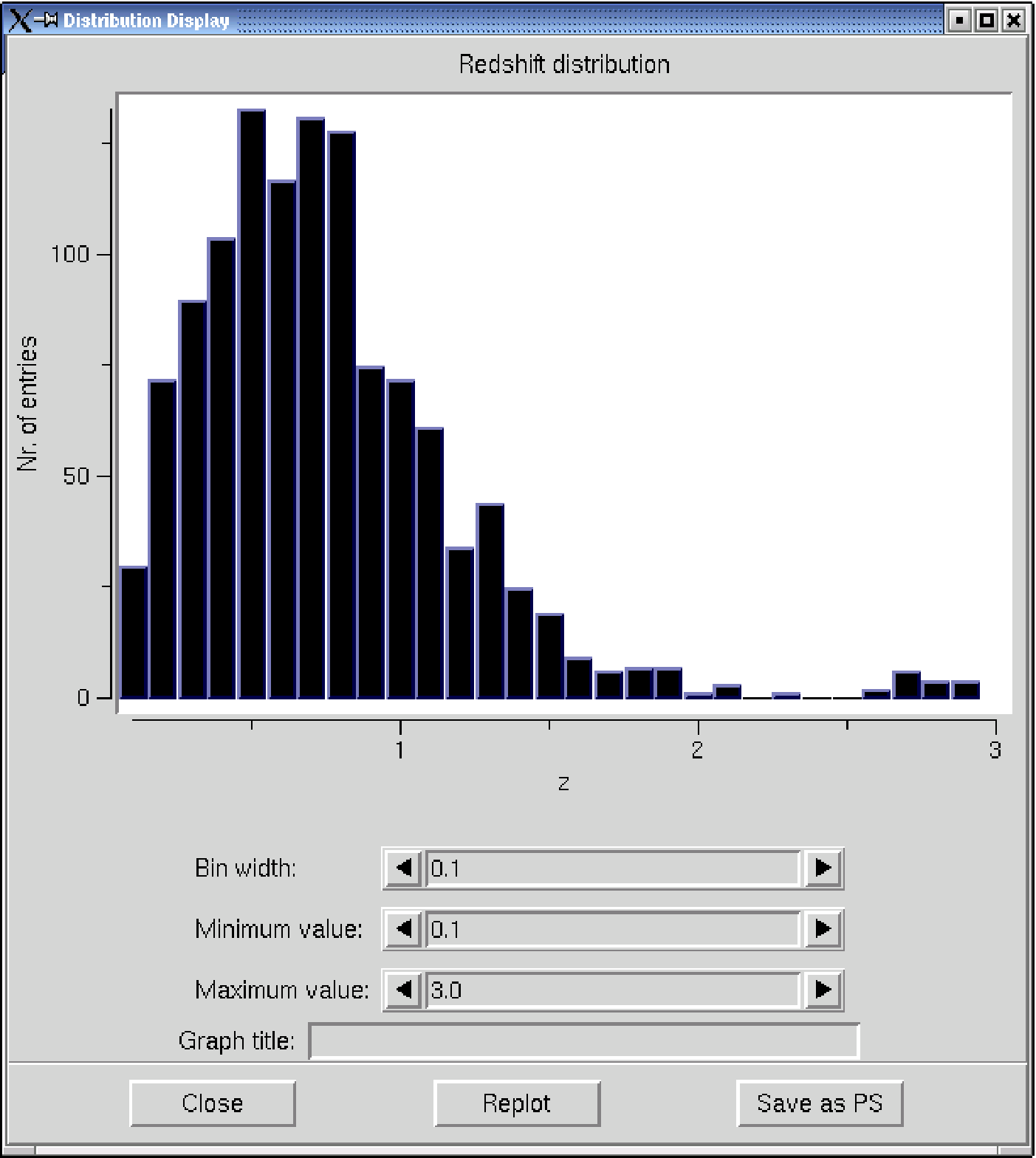}
  \end{tabular}
  \caption{DBrowser screenshots. An example of database inquiring and statistical plot. Here we have
  a redshift distribution ore a sky region defined in the query.}
  \label{figdbb}
\end{figure*}

It is designed to be a quick analysis tool for scientists who have their data stored
in tables within a MySQL database, as it provides simple and easy to use statistical
and plotting facilities: plot data value distributions or correlations; get measurements for a distribution
mean, median, and spread; measure the degree of correlation and derive the linear correlation
coefficients between two sets of data values (see Figure ~\ref{figdbb}).

Being embedded within DRAGO, it will eliminate the requirement of extracting the data from the
database and passing them to some external package for analysis.

Once incorporated in the DRAGO plug-in suite, it will be possible to complete the
DRAGO philosophy, having a global framework that satisfied the three requirements of
data organization, data reduction and analysis, data storing.

\end{document}